\documentclass[aps,prd,twocolumn,preprintnumbers,floatfix,nofootinbib]{revtex4}

\usepackage[usenames]{color}
\usepackage{subfigure,amsmath, amssymb ,amsfonts, latexsym}  
\usepackage[final]{graphicx}
\usepackage[colorlinks,hyperindex, 
            bookmarks=true,bookmarksopen=true]{hyperref}
    \hypersetup
    {
        colorlinks,%
        citecolor=red,%
        linkcolor=blue,%
        urlcolor=black,%
    }

\usepackage[notcite, color, final
                               ]{showkeys} 
\definecolor{refkey}{gray}{.5} 
\definecolor{labelkey}{gray}{.5}

       \def\G  {\Gamma}

\renewcommand{\a}{\alpha}      
      
\renewcommand{\l}{\lambda}

 \newcommand{\caln}{\mbox{${\cal N}$}}

\newcommand{\be}{\begin{equation}}
\newcommand{\ee}{\end{equation}}
\newcommand{\beqa}{\begin{subequations}\begin{eqnarray}}
\newcommand{\eeqa}{\end{eqnarray}\end{subequations}}
\newcommand{\nn}{\nonumber}

\newcommand{\ra}{\rightarrow}

\newcommand{\dd}{\mathrm{d}}

\newcommand{\tA}{{\tilde{A}}}

\begin{document}

\newcommand\sect[1]{\emph{#1}---}

\preprint{
\begin{minipage}[t]{3in}
\begin{flushright} SHEP-11-26
\\[30pt]
\hphantom{.}
\end{flushright}
\end{minipage}
}

\title{Towards a Holographic Model of the QCD Phase Diagram}

\author{Nick Evans}
\email{evans@soton.ac.uk}
\affiliation{ School of Physics and Astronomy, University of
Southampton, Southampton, SO17 1BJ, UK 
}
\author{Astrid Gebauer}
\email{ag806@soton.ac.uk}
\affiliation{ School of Physics and Astronomy, University of
Southampton, Southampton, SO17 1BJ, UK 
}
\author{Keun-Young Kim}
\email{K.Y.Kim@uva.nl}
\affiliation{ School of Physics and Astronomy, University of
Southampton, Southampton, SO17 1BJ, UK  
}
\affiliation{ Institute for Theoretical Physics, University of Amsterdam, Science Park 904, \\ Postbus 94485, 1090 GL Amsterdam, The Netherlands}
\author{Maria Magou}
\email{mm21g08@soton.ac.uk}
\affiliation{ School of Physics and Astronomy, University of
Southampton, Southampton, SO17 1BJ, UK  
}

\begin{abstract}
\noindent We describe the temperature-chemical potential 
phase diagrams of holographic models of a range of strongly coupled gauge theories
that display chiral symmetry breaking/restoration transitions. The models are based
on the D3/probe-D7 system but with a phenomenologically chosen running coupling/dilaton 
profile. We realize chiral phase transitions with either temperature or density that are first or
second order by changing the dilaton profile. Although
the models are only caricatures of QCD they show that holographic models can capture
many aspects of the QCD phase diagram and hint at the dependence on the running coupling.
  
\end{abstract}

\maketitle


\section{Introduction}

The QCD phase diagram is notoriously difficult to compute. Firstly the physics
associated with deconfinement or chiral symmetry restoration is strongly coupled
where we traditionally do not know how to compute. Secondly at finite density
lattice gauge theory, the first principles simulation of the theory on supercomputers,
suffers a ``sign problem'' that means Monte Carlo methods break down. In fact with light quarks there is no
clear order parameter for deconfinement so we will concentrate on the chiral transition. Progress
has been made by identifying effective theories of the transitions and through lattice computations 
at low density. 
\cite{Rajagopal:1999cp} provides a review of the standard picture. It is believed for
QCD, with the physical quark masses, that the phase transition with temperature is a smooth
cross over (becoming a second order transition as the up and down quark masses go to zero). 
At zero temperature the transition with density is believed to be first order. There must
therefore be a critical point where the first order line ends in the temperature density
plane.

In the last ten years the AdS/CFT Correspondence \cite{Maldacena:1997re,Gubser:1998bc,Witten:1998qj}
and more general gauge/gravity dualities have
emerged as a new tool for the study of strongly coupled problems.  It is interesting to ask
whether these holographic models can in principle describe a phase
diagram like that of real QCD. One recent attempt in this direction can be found in \cite{DeWolfe:2010he} 
where a 5d holographic model of a strongly coupled gauge theory with a running coupling was shown to give
an appropriate looking phase diagram. In that model the order parameter is associated with confinement  
and it does not include quark fields. In true QCD the order parameter across the phase diagram is the quark bilinear
condensate. In this paper we will study a holographic model of quark fields and again attempt to reproduce 
key features of the QCD phase diagram.

Quarks can be introduced into the AdS/CFT Correspondence through
probe branes \cite{Grana:2001xn,Bertolini:2001qa,Karch:2002sh,Kruczenski:2003be,Erdmenger:2007cm}
and a number of systems with chiral symmetry breaking have been developed \cite{Babington:2003vm,Kruczenski:2003uq,Ghoroku:2004sp,Alvares:2009hv,Filev:2007gb,Sakai:2004cn}.
We can not of course describe true QCD holographically because
the dual, if it exists, is not known and is probably very complicated (and strongly coupled, at least, in the UV). 
Our analysis is therefore in the
spirit of AdS/QCD \cite{Erlich:2005qh,Da Rold:2005zs}, a phenomenological modeling of the QCD phase diagram. If one could model the
phase diagram correctly one might hope to then predict other features of the theory such as time
dependent dynamics during transitions and so forth. 

\begin{figure}[]
\centering
   {\includegraphics[width=6cm]{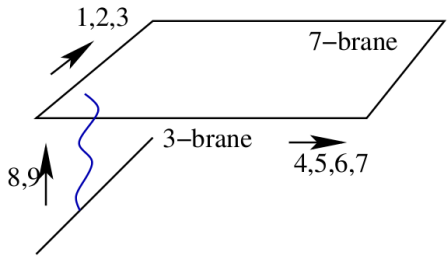}}
\caption{
           {A schematic of the D3/D7 showing our conventions. The D3-D3 strings generate the 
           ${\cal N}=4$ theory, the D3-D7 string 
           represent the quarks and D7-D7 strings describe mesonic operators. }
           }\label{73b}
\end{figure}

Our models will be in the context of the simplest brane construction of a 3+1d gauge theory with
quarks which is the D3/D7 system of Fig \ref{73b} \cite{Grana:2001xn,Bertolini:2001qa,Karch:2002sh,Kruczenski:2003be,Erdmenger:2007cm}. The basic gauge theory is large $N$, 
${\cal N}=4$ super Yang-Mills with $N_f$ quark fields. We will work in the quenched approximation
where we neglect quark loops. On the gauge theory side we do not backreact the D7 branes, that provide
the quarks, on the geometry but instead work in the probe approximation \cite{Karch:2002sh}. The theory
has a $U(1)$ symmetry under which a fermionic quark anti-quark condensate has charge 2 and plays the role
of $U(1)$ axial \cite{Babington:2003vm}. Although the theory does not have a non-abelian chiral symmetry (as for example the
Sakai Sugimoto model does \cite{Sakai:2004cn}) this is not important in the quenched approximation since the 
dynamics of the formation of the quark condensate is flavour independent. The model is very simple to work 
with having a background geometry that is just AdS$_5\times S^5$ and the gauge theory is 
3+1d at all energy scales.

So called top down models of this type exist with chiral symmetry breaking. Supergravity solutions exist that
correspond to the AdS space being deformed in reaction to a running coupling introduced by a
non-trivial dilaton profile \cite{Babington:2003vm,Ghoroku:2004sp}. In cases where the coupling grows in the infra-red (IR),
breaking the conformal symmetry,  chiral symmetry
breaking is induced. These models have very specific forms for the running coupling and
are typically singular somewhere in the interior. At the string theory level a full interpretation
is lacking. 

A yet simpler and completely computable case with chiral symmetry breaking
is provided by introducing a background magnetic field
associated with $U(1)$ baryon number \cite{Filev:2007gb}. Such a background source can be described by a gauge field on 
the surface of the D7 brane. A chiral condensate is induced. Very simplistically one can think
of the $B$-field as introducing a scale that breaks the conformal symmetry as the strong
coupling scale $\Lambda_{\rm QCD}$ does in QCD allowing the strong dynamics to form the quark condensate.

\begin{figure}[]
\centering
  \subfigure[$T$-$\mu$ phase diagram. The inset diagrams are representative  probe brane embeddings (dotted lines), 
    where a black disk represents a black hole.]
   {\includegraphics[width=7cm]{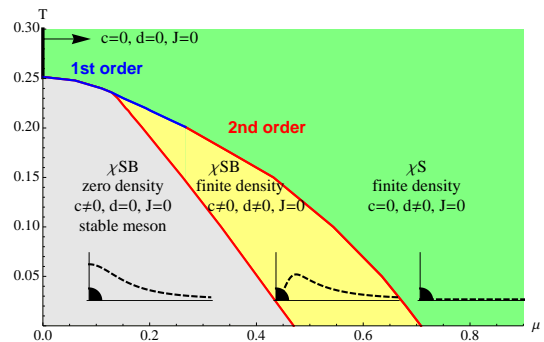} \label{Tvsmu}} \quad
  \subfigure[$T$-$E$ phase diagram. The inset diagrams are representative probe brane embeddings (dotted lines), 
    where a red arc represents a singular shell.]
   {\includegraphics[width=7cm]{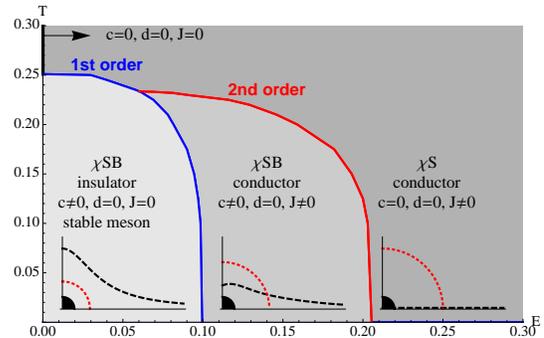} \label{TvsE} }
  \caption{{The phase diagrams of the massless ${\cal N}=2$ gauge theory with
  a magnetic field. First order transitions are shown in blue,
  second order transitions in red. The temperature is controlled by the parameter $T$,
  chemical potential by $\mu$ and electric field by $E$.}
           } \label{phase0}
\end{figure}

In recent papers \cite{Evans:2010iy,Evans:2010np,Evans:2011mu,Evans:2011tk,Evans:2010xs} we have explored the phase structure of the theory with magnetic field.
Temperature can be introduced through an AdS Schwarzschild black hole in the geometry \cite{Witten:1998qj}. Density and
chemical potential can be added through the temporal component of the $U(1)$ baryon number gauge
field \cite{Nakamura:2006xk,Kobayashi:2006sb,Kim:2006gp}. The phase diagram \cite{Evans:2010iy} is shown in Fig \ref{Tvsmu}. 
The chiral restoration transition was found to be first order with temperature and second
order with density. A critical point lies between these regimes. In addition there is an extra transition 
associated with the formation of density and the mesons of the system melting into the background plasma \cite{Peeters:2006iu,Hoyos:2006gb}.
In places in phase
space this merges with the chiral transition but in other places it separates and can be either first 
or second order. Such a phase with a quark density but chiral symmetry breaking could potentially exist in 
QCD. It would be nice to also describe this transition as a deconfinement transition
but firstly the ${\cal N}=4$ background does not induce linear confinement and secondly the presence
of any temperature leads to screening of the quarks at the length scale of the inverse temperature.
The meson bound states are closer in spirit to atomic bound states than QCD-like mesons. Nevertheless
they are being disrupted by the background plasma so the existence of a phase with melted mesons
but chiral symmetry breaking at least leads one to speculate on a possible separation of deconfinement and chiral
restoration behaviour in the QCD phase plane.

\begin{figure}[]
\centering
   {\includegraphics[width=7cm]{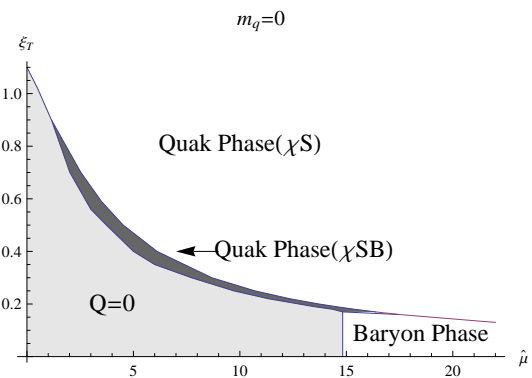} }  
  \caption{{The phase diagram of the massless axion/dilaton gauge theory in \cite{Gwak:2011wr}.  }
           }  \label{dilax}
\end{figure}

In the gravity picture of the transitions the crucial transitions are between the three sorts of D7 
brane embedding shown in Fig \ref{Tvsmu}. An embedding that curves in the holographic space
to miss the black hole represents the chiral symmetry breaking phase. These solutions have stable
and discrete linearized fluctuations that correspond to the mesons of the theory. The second type of
embedding 
curves off axis but ends on the black hole. It describes a phase with chiral symmetry breaking
but the linearized fluctuations are now replaced by in-falling quasi-normal modes of the
black hole which describe unstable mesonic fluctuations of a quark plasma.
Finally the flat embedding ending on the black hole describes melted mesons and chiral symmetry restoration.
The transitions occur when the  magnitude of the D7 action for these 
different cases cross.

How generic to the holographic description is this phase structure? Keeping within the top down analysis one
can change parameters and see what effect they have on the phase diagram. For example in \cite{Evans:2011mu}
 we
traded density for an electric field parallel to the magnetic field.  The electric field tries to dissociate
mesons by accelerating the quark and anti-quark in opposite directions and so opposes the formation
of a chiral condensate. The phase diagram of the theory in the $T$-$E$ plane at fixed $B$ is also
shown in Fig \ref{TvsE}. The chiral transitions are again first order with temperature but second order with density
although there is some change in the meson melting transition order\footnote{There may be an instability in Fig \ref{phase0} due to the WZ term contribution \cite{Evans:2011tk,Kharzeev:2011rw}.}. We followed up this analysis in
\cite{Evans:2011tk} to consider the case of mutually perpendicular electric and magnetic fields - there the chiral transition
is first order in nature throughout the full $T$-$E$ plane. 

A recent paper \cite{Gwak:2011wr} performed a simliar analysis with a running dilaton geometry. The geometry is
that of \cite{Ghoroku:2005tf} in which there is  a non-zero profile for both the dilaton and axion
fields in AdS. The field theory is the ${\cal N}=4$ gauge theory with a vev for both Tr$F^2$ and Tr$F \tilde{F}$
which preserves supersymmetry at zero temperature but displays confinement. A D7 was introduced in a
supersymmetry breaking fashion and chiral symmetry breaking is observed. The temperature density chiral
transformation was first order throughout the plane and is shown in Fig \ref{dilax}. It shows the same three
phases as the magnetic field case. An extra component of the analysis in \cite{Gwak:2011wr} was to note that in the confining geometry with a running
dilaton a baryonic phase was also present. A baryon vertex is described by a D5 brane wrapped on the 
$S^5$ of the AdS$_5 \times S^5$ space. In the pure ${\cal N}=4$ theory such vertices shrink to zero. However in the 
running dilaton geometry the large IR value of the dilaton stabilizes the D5 embedding. Solutions
exist that link the D5 to the D7 brane embedding with a balancing force condition. These configurations
describe the gauge theory with finite baryon density rather than finite quark density. This phase
sets in at finite chemical potential and then persists to infinite chemical potential (as
shown in Fig \ref{dilax}) which is certainly unlike QCD. We will not focus on this phase in this paper
but it would be interesting to study it in future work to find models that have a baryonic phase in
some intermediate range of chemical potential like QCD.

These phase structures are very interesting and surprisingly complex but do not match the expectations
in QCD. In QCD we need a second order transition with temperature and a first order transition with density
to the chirally symmetric phase.

Here we want to work in a much more generic framework to ask what phase structures it
is possible to get in the holographic description and to try to force ourselves onto a 
representation of the QCD phase diagram. We will therefore take a bottom up approach within
the model and allow ourselves to dial the running of the gauge coupling by hand. We will have a dilaton
profile that smoothly transitions from a UV conformal regime to an IR conformal regime 
through a step of variable height and width. Such an ansatz allows one to consider
runnings that range from precocious growth in the IR to more walking like dynamics\cite{Holdom:1981rm}. We used a similar
ansatz in previous papers to study the impact of walking \cite{Alvares:2009hv} on meson physics and as a mechanism
for generating inflation \cite{Evans:2010tf}. Here, in a completely new analysis of
the phase structure of these models, we find that with the simple step ansatz we can move from a totally 
first order transition in the phase plane to a configuration similar to that we obtained
with a $B$ field (a first order transition with temperature but second order with density). With this ansatz we can not achieve a second order transition with temperature.

The model directly suggests other phenomenological generalizations though. In particular, if
we think of the running dilaton profile as a short cut for including the back reaction
due to the quark fields/D7 brane \footnote {In \cite{Bigazzi:2011it}, D3-D7 solutions at finite temperature and chemical potential, with the inclusion of dynamical flavor effects, have been derived and studied as 
full-fledged top-down models. 
They will be the first step towards the top-down study of phase transitions in D3-D7 systems with dynamical flavors.}, then it is natural to break the $SO(6)$ symmetry of AdS$_5$ in the dilaton
in the same fashion as the D3/D7 system's geometry. This allows us an extra phenomenological
freedom to distort the dilaton or black hole horizon. These simple changes do allow us to 
reproduce a wide range of phase diagrams including QCD-like ones as we will show below. We will discuss the
simple geometric reasons for the emergence of first or second order transitions in these different scenarios. 

Our conclusion therefore is that
the holographic model has no intrinsic problem with mimicking the QCD phase diagram and these systems
may therefore be phenomenologically useful in the future.

\section{The Model}

First let us review the gravity dual description of the symmetry
breaking behaviour of our strongly coupled gauge theory. 

Dp-branes are p dimensional membrane like objects to which the
ends of open strings are tied. The weak coupling picture for our D3/D7
set up is shown in Fig \ref{73b} \cite{Grana:2001xn,Bertolini:2001qa,Karch:2002sh,Kruczenski:2003be,Erdmenger:2007cm} - there are $N$ D3 branes and the
lightest string states with both ends on the D3 generate the
adjoint representation fields of the ${\cal N}=4$ gauge theory.
Strings stretched between the D3 and the D7 are the quark fields
lying in the fundamental representation of the $SU(N)$ group (they
have just one end on the D3).

In the strong coupling limit the D3 branes in this picture are
replaced by the geometry that they induce. We will consider a
gauge theory with a holographic dual described by the Einstein
frame geometry AdS$_5 \times S^5$ 
\begin{equation} 
\dd s^2 = 
{r^2 \over R^2} \dd x_{4}^2 + {R^2 \over r^2} \left( \dd \varrho^2 +
\varrho^2 \dd \Omega_3^2 + \dd w_5^2 + \dd w_6^2 \right) \,,
\end{equation}
where we
have split the coordinates into the $x_{3+1}$ of the gauge theory,
the $\varrho$ and $\Omega_3$ which will be on the D7 brane
world-volume and two directions transverse to the D7, $w_5,w_6$.
The radial coordinate, $r^2 = \varrho^2 + w_5^2 + w_6^2$, corresponds
to the energy scale of the gauge theory. The radius of curvature
is given by $R^4 = 4 \pi g_s N \alpha^{'2}$ with $N$ the
number of colours.  The $r \rightarrow \infty$ limit of this
theory is dual to the ${\cal N}=4$ super Yang-Mills theory where 
$g_s = g^2_{\mathrm{UV}}$ is the constant large $r$ asymptotic value of the gauge
coupling.

In addition we will allow ourselves to choose the profile of the dilaton as
$r \rightarrow 0$ to represent the running of the gauge theory coupling, 
%
%
$e^\phi \equiv \beta$,
where the function $\beta \rightarrow 1$ as
$r \rightarrow \infty$.
An interesting phenomenological case  is to consider a gauge
coupling running with a step of the form \cite{Alvares:2009hv,Evans:2010tf}
\begin{equation} \label{coup}
 \beta(r) =  A + 1 - A \tanh\left[ \Gamma (r - \lambda) 
\right] \,. \end{equation}
Of course in this case the geometry is not back
reacted to the dilaton and the model is a phenomenological one in
the spirit of AdS/QCD\cite{Erlich:2005qh,Da Rold:2005zs}. This form introduces conformal symmetry
breaking at the scale $\Lambda = \lambda/2 \pi \alpha'$ which
triggers chiral symmetry breaking. The parameter $A$ determines
the increase in the coupling across the step.

We will introduce a single D7 brane probe \cite{Karch:2002sh} into the
geometry to include quarks - by treating the D7 as a probe we are
working in a quenched approximation although we can reintroduce
some aspects of quark loops through the running coupling's form if
we wish (or know how). This system has a $U(1)$ axial symmetry on
the quarks, corresponding to rotations in the $w_5$-$w_6$ plane,
which will be broken by the formation of a quark condensate.

In the true vacuum at $T=0$ the brane will be static. We must find
the D7 embedding function e.g. $w_5(\varrho), w_6=0$. The Dirac Born
Infeld (DBI) action in Einstein frame is given by 
\begin{equation}
\begin{split}
S_{D7}  &=  -T_7 \int \dd^8\xi e^\phi  \sqrt{- \det P[G]_{ab}} \\
 &=  -\overline{T}_7 \int \dd^4x~ d \varrho ~ \varrho^3 \beta \sqrt{1 +
(\partial_\varrho w_5)^2} \,,
\end{split}
\end{equation} 
where $T_7 = (2 \pi)^{-7} \alpha'^{-4}g^{-2}_{\mathrm{UV}}$ and $\overline{T}_7 = 2 \pi^2 T_7$ when we
have integrated over the 3-sphere on the D7. The equation of
motion for the embedding function is therefore 
\begin{equation} \label{embed}
\partial_\rho \left[ {\beta \varrho^3
\partial_\varrho w_5 \over \sqrt{1+ (\partial_\varrho w_5)^2}}\right] - 2 w_5 \varrho^3
\sqrt{1+ (\partial_\varrho w_5)^2} {\partial \beta \over \partial
r^2} = 0 \,. 
\end{equation}
The UV asymptotic of this equation, provided the
dilaton returns to a constant so the UV dual is the ${\cal N}=4$
super Yang-Mills theory, has solutions of the form 
\begin{equation}
\label{asy}w_5 = m + {c \over \varrho^2} + \cdots \,,
\end{equation}
where we can
interpret $m$ as the quark mass ($m_q = m/2 \pi \alpha'$) and $c$
is proportional to the quark condensate.

The embedding equation (\ref{embed}) clearly has regular solutions
$w_5=m$ when $g^2_{YM}$ is independent of $r$ - the flat
embeddings of the ${\cal N}=2$ Karch-Katz theory \cite{Karch:2002sh}.
Equally clearly if $\partial \beta / \partial r^2$ is none trivial
in $w_5$ then the second term in (\ref{embed}) will not vanish for
a flat embedding.

There is always a solution $w_5=0$ which corresponds to a massless
quark with zero quark condensate ($c=0$). In the pure $\caln=2$ gauge theory 
with $\beta=1$ this
is the true vacuum. In the symmetry breaking geometries this
configuration is a local maximum of the potential.

If the coupling is
larger near the origin then the D7 brane will be repelled
from the origin \footnote{In fact there is a competition between
the increased action from the D7 entering the region with larger dilaton
and the derivative cost of the D7 bending to avoid it. This leads to a critical value of
$A$ to trigger chiral symmetry breaking. For example for $\lambda=1.7$ and $\Gamma=1$ $A_c=2.1$.
In this paper we will consider only super-critical values of $A$.}.  The parameter $\Gamma$ spreads the increase in
the coupling over a region in $r$ of order $\Gamma^{-1}$ in size.

\begin{figure}[]
\centering
    {\includegraphics[width=7cm]{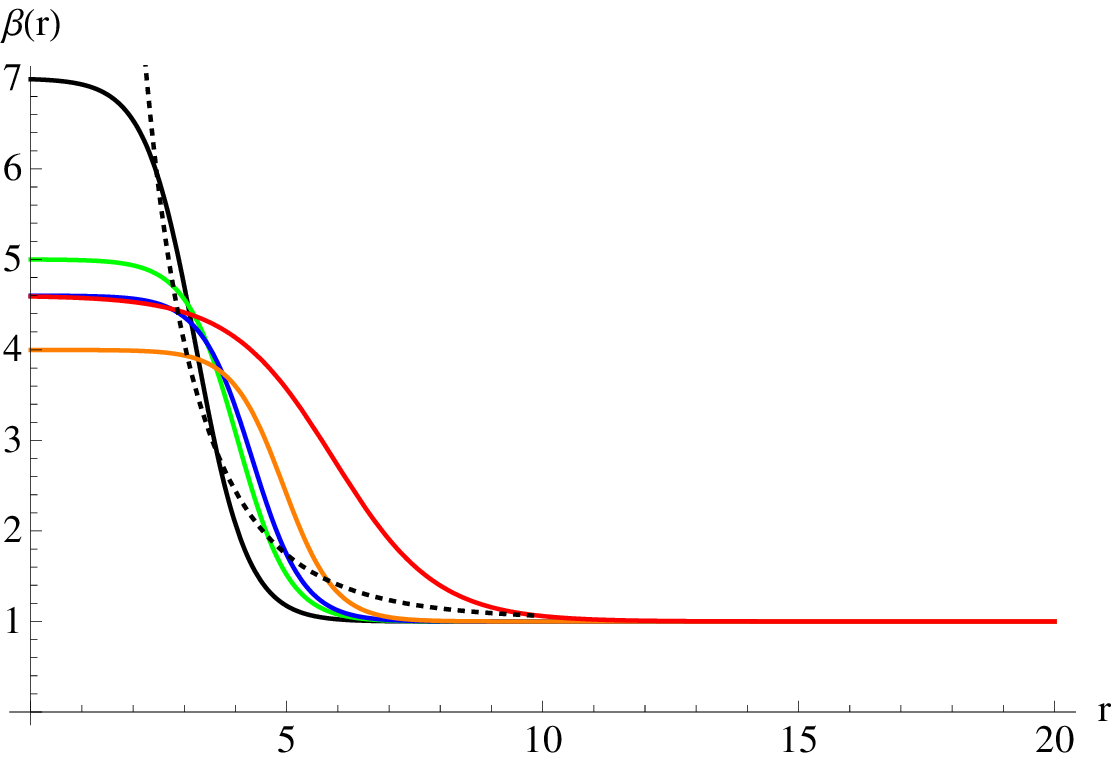}}
    {\includegraphics[width=7cm]{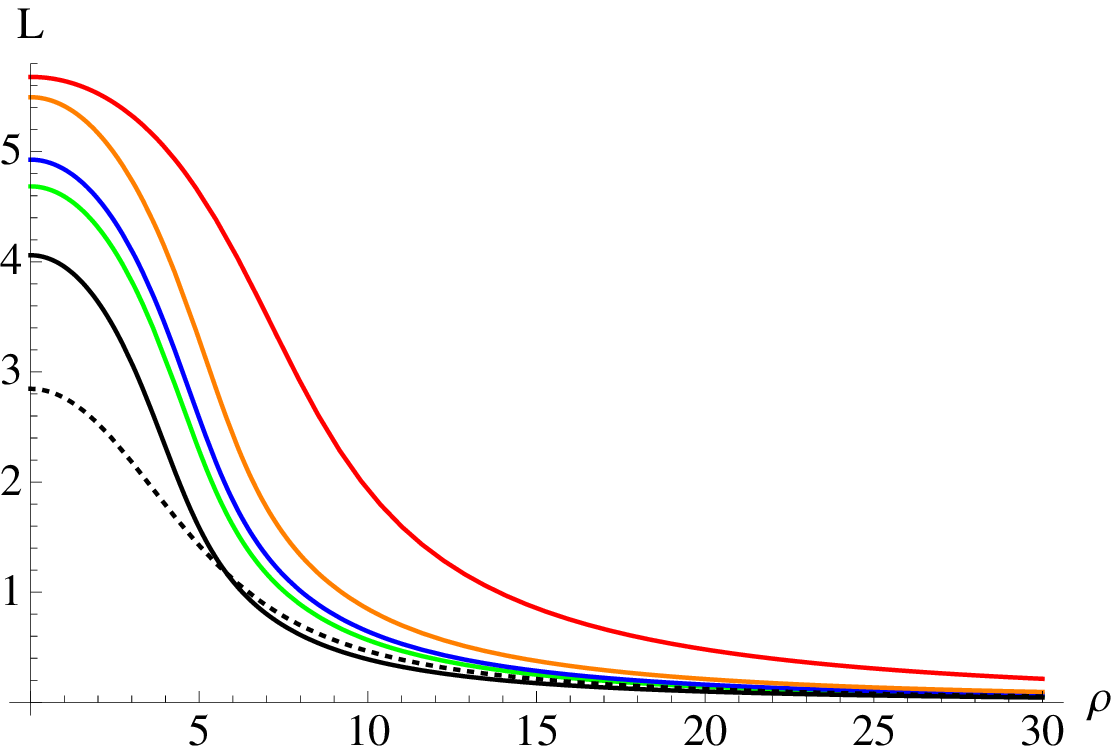}}
      \begin{tabular}{|c|c|c|c|}
      \hline  & $\G$ & $A$ & $\l$ \\
      \hline Black & 1 & 3 & 3.240 \\
      \hline Green & 1 & 2 & 4.045 \\
      \hline Blue & 1 & 1.8 & 4.325 \\
      \hline Orange & 1 & 1.5 & 4.940 \\
      \hline Red & 0.5 & 1.8 & 5.882 \\
      \hline Black(Dotted) & B = 35.6 & - & - \\
      \hline
   \end{tabular}
       \caption{
           { Example coupling flows (\ref{coup}) (top) and 
            the induced D7 brane embeddings/quark self energy (bottom) with the parameter choices shown in the table.
           }
           }\label{Embed}
\end{figure}

We display the embeddings for some particular cases in Fig
\ref{Embed}. Note that we have chosen parameters here that make the vacuum
energy of the theory the same in each case. The vacuum energy is
given by minus the DBI action evaluated on the solution. In fact this
energy is formally divergent corresponding to the usual
cosmological constant problem in field theory. As usual we will
subtract the UV component of the energy to renormalize.

The symmetry breaking of these solutions is visible directly \cite{Babington:2003vm}. The $U(1)$
symmetry corresponds to rotations of the solution in the $w_5$-$w_6$
plane. An embedding along the $\varrho$ axis corresponds to a massless quark
with the symmetry unbroken (this is the configuration that is
preferred at high temperature and it has zero condensate $c$).
The symmetry breaking configurations though map onto the flat case at large $\varrho$
(the UV of the theory) but bend off axis breaking the symmetry in
the IR.

One can interpret the D7 embedding function as the dynamical self
energy of the quark, similar to that emerging from a gap equation \cite{Alvares:2009hv}.
The separation of the D7 from the $\varrho$ axis is the mass at some
particular energy scale given by $\varrho$.

\subsection{Temperature}

Temperature can be included in the theory by using the AdS-Schwarzschild 
black hole metric. 
In Einstein frame we have
\begin{equation} 
\dd s^2 =  -{K(r) \over R^2} \dd t^2 + {R^2 \over K(r)}\dd r^2+ \frac{r^2}{R^2} \dd\vec{x}_3^2
+ R^2 d\Omega_5^2 \,, 
\end{equation}
where
\begin{equation}
 K(r) = r^2-\frac{r_H^4}{r^2} \,, \quad \qquad r_H = \pi R^2 T \,.
\end{equation}

Witten identified this as the thermal description of the gauge theory in \cite{Witten:1998qj}.
The parameter $r_H$ is of dimension one in the field theory and preserves
the $SO(6)$ symmetry so is identified as shown with temperature T. The black hole
is the natural candidate since it has intrinsic thermodynamic properties such
as entropy and temperature.

We make the coordinate transformation \cite{Babington:2003vm} 
\begin{equation}
\begin{split}
  &\frac{rdr}{(r^4-r_H^4)^{1/2}} \equiv \frac{dw}{w} \,, \\
  &2w^2 = r^2 + \sqrt{r^4 - r_H^4}\ , 
\end{split}
\end{equation}
with $\sqrt{2}w_H = r_H$, such that the metric becomes
\begin{equation} 
\begin{split}
   \dd s^2 &=   \frac{w^2}{R^2}(- g_t \dd t^2 + g_x \dd\vec{x}^2)  \\
   & + \frac{R^2}{w^2} (\dd\rho^2 + \rho^2 \dd\Omega_3^2
         + \dd L^2 + L^2 \dd\Omega_1^2 ), 
\end{split} 
\end{equation}
where
\begin{equation}
\begin{split}
& g_t = \frac{(w^4 - w_H^4)^2}{ w^4 (w^4+w_H^4)}\,,  \quad
g_x  = \frac{w^4 + w_H^4}{ w^4} \,, \\
& w = \sqrt{\rho^2 + L^2}\,,  \quad \rho = w \sin\theta \,,
  \quad L = w \cos\theta \,,
\end{split}
\end{equation}

Now we have to transform $\beta$ also:
\begin{equation} 
e^{\phi} = \beta \left(\frac{w^4+w_H^4}{w^2}\right) \,,
\end{equation}
and therefore
\begin{equation}
\beta =  A + 1 - A \tanh\left[
\Gamma \left(\sqrt{\frac{(\rho^2 + L^2)^2+w_H^4}{\rho^2 + L^2}} - \lambda\right) \right]\,. \label{beta}
\end{equation}
Note that for $w_H \rightarrow 0$:  $w \ra r$, $\rho \ra \varrho$ and $L \ra w_5$, if we set $w_6 = 0$. 

\subsection{Chemical Potential}

The DBI action for the D7 brane naturally includes a surface gauge field which
holographically describes the quark bilinear operators $\bar{q} \gamma^\mu q$ 
and their source, a background $U(1)$ baryon number gauge field \cite{Nakamura:2006xk,Kobayashi:2006sb,Kim:2006gp}.
We introduce a chemical potential through the $U(1)$ baryon number gauge field 
which enters the DBI action in Einstein frame  as 
\begin{equation}
{S_{D7}  =  -T_7 \int \dd^8\xi e^{-\phi}  \sqrt{- \det \left(e^{\phi/2}P[G]_{ab}+(2 \pi \alpha')F_{ab} \right)} \,. \nn}
\end{equation}

We allow a chemical potential through $A_t(\rho) \neq 0$. So the Action becomes
\begin{equation} 
\begin{split}
S_{D7}&=\int d^4x\ d\rho\ \mathcal{L} \\ 
&=-\overline{T}_7\int d^4x\ d\rho\ \beta (\rho) \rho^3\sqrt{g_t g_x^3\left(1+L'^2\right)-  \frac{g_x^3}{\beta(\rho)}  \tA_t'^2}  \,. 
\end{split}
\end{equation}
where  $\tA_t'=(2\pi\a')A_t'$.
In our convention of the metric this is
\begin{equation}
\begin{split}
\mathcal{L} & =  -\overline{T}_7 \beta (\rho) \rho^3\ \left( 1-\frac{w_H^4}{w^4}\right) \left( 1+\frac{w_H^4}{w^4}  \right) \\
&  \sqrt{\left(1+L'^2\right)- \frac{w^4 \left( w^4 +w_H^4\right) }{\left(w^4 -w_H^4 \right)^2 }  \frac{ \tA_t'^2}{\beta(\rho)} }  \,. 
\end{split}
\end{equation}
Now we can Legendre transform the action as we have a conserved quantity, the density, $d \left(= \frac{\delta S_{D7}}{\delta A_t'}\right)$.
\begin{equation}
\tilde{S}_{D7}\ =\ S_{D7}-\int \dd^8\xi A_t' \frac{\delta S_{D7}}{\delta A_t'}\ =\  \left( \int_{S3} \epsilon_3 \int \dd^4x\right)\int \dd\rho\ \tilde{\mathcal{L}} \,, \nn
\end{equation}
where
\begin{equation} 
\begin{split}
\tilde{\mathcal{L}} & =  -\overline{T}_7 \frac{\left(w^4-w_H^4\right)}{\left( w ^4\right)} \sqrt{1 + L^{'2}} \\
& \sqrt{\left(\frac{ w ^4d^2 \beta(\rho)}{\left((2 \pi \alpha' )^2 \overline{T}_7^2 \left(w ^4+w_H^4\right)\right)}+\frac{ \rho ^6 \left(w ^4+w_H^4\right)^2}{ w ^8}\beta(\rho)^2\right)}. 
\end{split} 
\end{equation}

We can redefine $d = (2 \pi \alpha' )\overline{T}_7\tilde{d}$ to give the simpler expression
\begin{equation} 
\begin{split}
\tilde{\mathcal{L}} & = -\overline{T}_7 \frac{\left(w ^4-w_H^4\right)}{\left( w ^4\right)}\sqrt{1+L'^2} \\ 
& \sqrt{\left(\frac{ w ^4 \tilde{d}^2 \beta(\rho)}{ \left(w ^4+w_H^4\right))}+\frac{ \rho ^6 \left(w ^4+w_H^4\right)^2}{ w ^8}\beta(\rho)^2\right)} \,.
\end{split} \label{lagfin}
\end{equation}
By varying the Lagrangian with respect to $A_t'$, we get an expression for $\tilde{d} (\tilde{A}_t')$ which we can invert for  an expression for $\tilde{A}_t'(\tilde{d})$
\begin{equation} 
\begin{split}
 \tA_t' & = \tilde{d} \frac{\left(w ^4-w_H^4\right)}{\left(w ^4+w_H^4\right)} \sqrt{1+L'^2} \\ 
& \sqrt{\frac{ 1 }{ \frac{\tilde{d}^2}{\beta(\rho)}\frac{w ^4}{\left(w ^4+w_H^4\right)}+\rho ^6   \left(\frac{w ^4+w_H^4}{w ^4}\right)^2 }} \,.
\end{split}
\end{equation}
This can be used to find the chemical potential $\mu=\frac{\tilde{\mu}}{(2 \pi \alpha' )}$
\begin{equation} \label{mud}
\tilde{\mu}=\int_{\rho_H}^{\infty} \dd \rho \, \tilde{d} \frac{\left(w ^4-w_H^4\right)}{\left(w ^4+w_H^4\right)}\sqrt{\frac{ \left(1+L'^2\right) }{ \frac{\tilde{d}^2}{\beta(\rho)}\frac{w ^4}{\left(w ^4+w_H^4\right)}+\rho ^6   \left(\frac{w ^4+w_H^4}{w ^4}\right)^2 }}\,,
\end{equation}
where $\tilde{\mu}(\rho \rightarrow \rho_H)=0$.

The free energy can be found by integrating the Legendre transformed Lagrangian, the grand potential by integrating the original Lagrangian, where we replace $\tA_t'(d)$.
\begin{equation} 
\begin{split}
F  = -\frac{\tilde{S}_{D7}}{\overline{T}_7} &= \int_{\rho_H}^\infty \dd\rho\ \frac{\left(w ^4-w_H^4\right)}{\left( w ^4\right)} \beta(\rho) \sqrt{1 + L'^2} \nn \\ 
& \sqrt{ \frac{\tilde{d}^2}{\beta(\rho)}\frac{ w ^4 }{ \left(w ^4+w_H^4\right)}+\frac{ \rho ^6 \left(w ^4+w_H^4\right)^2}{ w ^8}}\,. 
\end{split} \nn 
\end{equation}
The grand potential is 
\begin{equation} 
\begin{split}
\Omega =-\frac{S_{D7}}{\overline{T}_7}&=\int_{\rho_H}^\infty \dd\rho\ \beta (\rho) \frac{w ^4-w_H^4}{w ^4} 
\rho ^{6}\left(\frac{w ^4+w_H^4}{w ^4}\right)^2 \nn \\ 
& \sqrt{\frac{ \left(1+L'^2\right)}{ \frac{\tilde{d}^2}{\beta(\rho)}  \frac{w ^4}{\left(w ^4+w_H^4\right)} +  \rho^6 \left(\frac{w ^4+w_H^4}{w ^4}\right)^2  }} \,, 
\end{split} \nn 
\end{equation}
where we need to note that $F(\rho \rightarrow \infty)=\Omega(\rho \rightarrow \infty)=\rho^3$, so we need to subtract $\frac{1}{4}(\Lambda_{UV})^4$ from both integrals to renormalize them.

\section{Analysis and results}

\begin{figure}[!h]
\centering
\subfigure[$T=0.4, A=30, \Gamma=0.1, \lambda=1.715$ - shows a single first order transition.]
   {{\includegraphics[width=5cm]{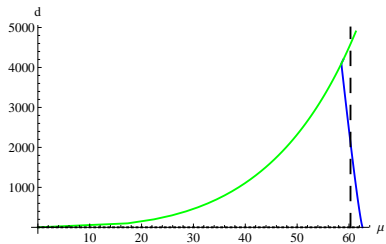}}} \qquad
\subfigure[$T=0.3, A=3, \Gamma=1, \lambda=1.715, \tilde{\alpha}=3$ - shows a first order transition followed by a second order transition as $\mu$ increases.]
   {\includegraphics[width=5cm]{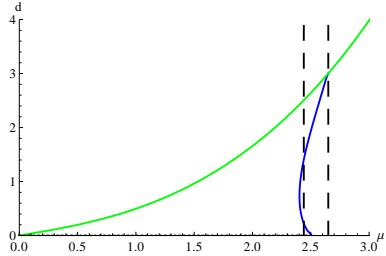}}
\subfigure[ $T=1.55, A=30, \Gamma=1, \lambda=1.715$ - shows a second order transition followed by a first order transition as $\mu$ increases.]
   {\includegraphics[width=5cm]{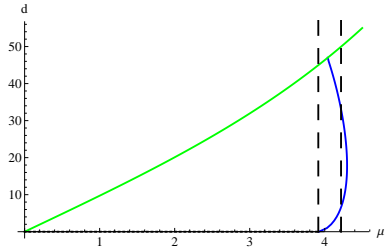}} \qquad
\subfigure[$T=0.7, A=3, \Gamma=1, \lambda=1.715, \tilde{\alpha}=3$ - shows two second order transitions.]
      {\includegraphics[width=5cm]{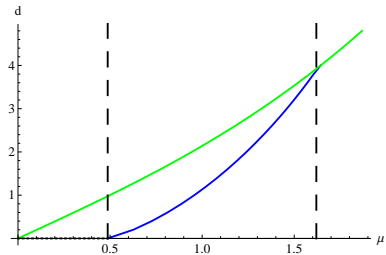}}
\caption{
           {Plots of density $d$ versus chemical potential $\mu$.  The top lighter line (green) in each case corresponds to the flat embedding; the horizontal line (black) along the axis is a chiral symmetry breaking (Minkowski) embedding; the near vertical dark line (blue) is a black hole embedding. Transition points are shown by the dotted vertical lines.}
           }\label{dmuplots}
\end{figure}

The methodology to study the phase diagram of our model is straight-forward if laborious.
We will work throughout in the massless quark limit.
We can think of the scale $\lambda$ in the dilaton ansatz as our intrinsic scale of
the theory and so we will leave that fixed. Then for each choice of parameters in
the dilaton profile ($A,\Gamma$) we analyze the theory on a grid in $T$ and $\mu$ space.

For each point on the $T, \mu$ grid we seek three sorts of embedding. The flat embedding $L=0$ exists in
all cases and describes the theory with $m=0$ and $c=0$. 
We use (\ref{mud}) to compute the $d$-$\mu$ relation for these embeddings.

We can also seek curved embeddings that miss the black hole. These solutions must have
$d=0$ but are consistent for any value of $\mu$. Here we use the equation of motion for $L$
from (\ref{lagfin}) and numerically shoot from an initial condition at $\rho=0$ with vanishing $\rho$ derivative, $L'(0)=0$.
We seek solutions that approach $L=0$ at large $\rho$. These configurations have a non-zero condensate parameter
$c$.  

Finally we can look for solutions that end on the black hole horizon. To find these we fix the density $d$ and 
shoot out from all points along the horizon seeking a solution that approaches $L=0$ at large $\rho$.
We then use (\ref{mud}) to compute $\mu$ from the solution. In this way we can fill out the $T$,$\mu$ grid.
The condensate can again be extracted from the large $\rho$ asymptotics of the embedding.

After finding as many such solutions as exist at each point the easiest method
to identify the transition points is to plot the density against $\mu$ on fixed $T$ lines. 
The transitions and their order are then manifest. We display four sample plots in Fig \ref{dmuplots} 
taken from scenarios below showing the four cases of the chiral transition and 
the meson melting transition being respectively first or second order in all combinations.

\subsection{Dependence on the change in coupling}

\begin{figure}[]
\centering
\subfigure[$A=3$   ]   
   {\includegraphics[width=5cm]{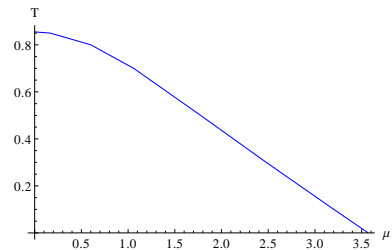}}
\subfigure[$A=15$  ]   
   {\includegraphics[width=5cm]{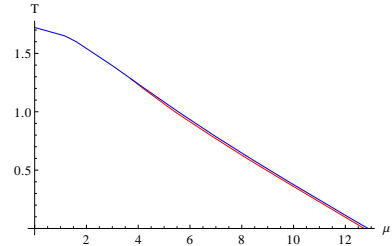}}
\subfigure[$A=30$  ]   
   {\includegraphics[width=5cm]{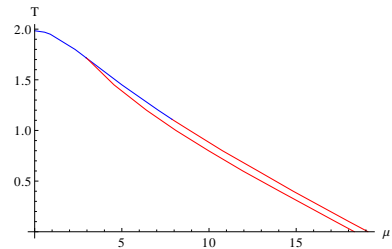}}   
  \caption{ Plots for three possible phase diagrams for the choices $A=3,15,30$.
    Large (small) $A$ gives second (first) order transition
    at low $T$.
    $\Gamma=1, \lambda = 1.715$.
           }\label{fig.a}
\end{figure}

Let us first consider how the phase diagram depends on the height of the step in
the gauge coupling function $\beta$. We fix $\lambda$ (the intrinsic scale of the theory)
and also $\Gamma=1$ and explore the phase structure as a function of $A$. We display the
results for three choices of $A$ in Fig \ref{fig.a}. 

In these and all our future phase diagrams the regions shown are 
similar to those in Fig \ref{Tvsmu} we will simply display the phase boundaries 
and their order henceforth. 

As mentioned in footnote 2 above there is a critical value of $A$ for chiral symmetry breaking
to occur. A conformal theory can not break a symmetry since it offers no scale for that symmetry
breaking to occur at. In fact some finite departure from conformality is needed to break
the chiral symmetry. 
For these choices of $\lambda, \Gamma$ the critical $A$ is $A_c=2.1$. We work above this value
throughout. 

At low $A$ there is a single transition for chiral symmetry restoration
and meson melting which is first order for all T and $\mu$. On the gravity side this is
a transition between the curved embedding that misses the black hole and the flat embedding.
In this case an embedding
ending on the black hole never plays a role.

For larger $A$, a new phase with chiral symmetry breaking but melted mesons develops. There is a regime
now in which the curved embedding ending on the black hole is energetically favoured. The transition
from the chiral symmetry breaking phase to this new phase is second order. The chiral symmetry restoration
phase remains first order.

At very large $A$ the chiral restoration transition becomes second order at high density. This latter phase 
resembles that of the theory with chiral symmetry breaking induced by a magnetic field \cite{Filev:2007gb}. 
In fact the B field case can be thought of as our case but with a choice of $\beta$ given by
\begin{equation}
\beta = \sqrt{1 + {B^2 w^4 \over (w^4 + w_H^4)^2}} \,.
\end{equation}

It is the black dotted curve ($w_H=0$) in Fig \ref{Embed} - it is not surprising therefore that we see similar phase structure here
(and indeed that we do provides strength to our analysis which is capturing  the behaviour
of top down models).

For very large $A$ the step becomes very sharp and there is little change relative to our phase diagram
in Fig \ref{fig.a}. In particular the thermal transition always remains first order.

\begin{figure}[]
\centering
\subfigure[$T=0.1$ ]   
   {\includegraphics[width=5cm]{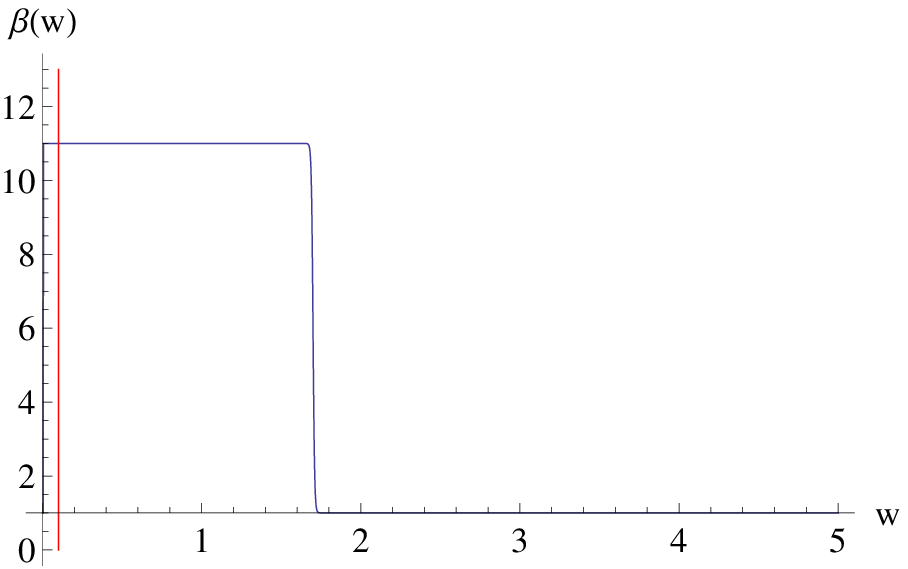}}
\subfigure[$T=1.0$ ]   
   {\includegraphics[width=5cm]{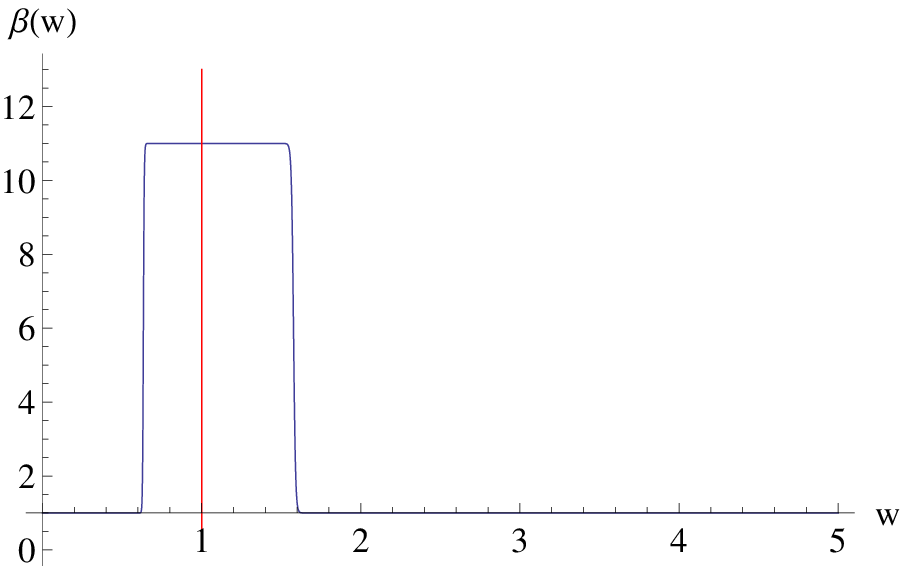}}
\subfigure[$T=1.20208 = \frac{\l}{\sqrt{2}}$. 
Dilaton is almost screened. The residual 
effect is due to finite $\G$ effect. ]   
   {\includegraphics[width=5cm]{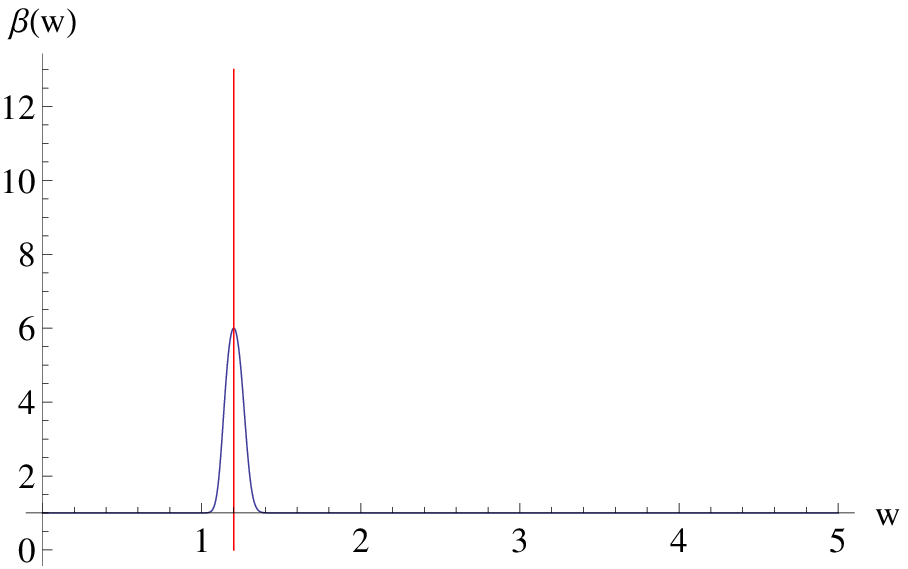}}   
  \caption{Plots for parameter choices
    $A=5, \Gamma=100, \lambda = 1.7$. The blue lines show the value of the coupling $\beta$. 
    The red line shows the position of the horizon. The final plot corresponds to the point of
    the first order transition.
    }\label{eating}
\end{figure}

The behaviour we are seeing here can be readily explained from the D7 perspective. First of all
the zero density transition with temperature is first order for a simple reason. The D7 embedding
breaks chiral symmetry at zero temperature because it
prefers to avoid the action cost of entering the region in which the dilaton is large. As temperature
is introduced through a small horizon the interior of the space is ``eaten'' but the D7 embedding
remains oblivious to this change since it never reaches down to small $r$. As temperature rises
the point of transition is when the 
horizon moves through the scale $\lambda$ where the dilaton step is. Once the region with a large
dilaton is eaten by the black hole the preferred D7 embedding is
the flat one. 

In Fig \ref{eating} we show an extreme case of this behaviour explicitly. Here we have taken $\Gamma$ very large so
that the transition in the dilaton between the low and high value is very sharp.   We plot the $\beta$ profile
against our radial parameter $w$ and mark in red the position of the black hole horizon. Note that in the 
$w$ coordinates the region where $\beta$ is large depends on the temperature (it doesn't in the 
orignal $r$ coordinate).
The dilaton effective radius $\l_*$ is 
\begin{equation}
  \l_* = \sqrt{\frac{\l^2+\sqrt{\l^4-4T^4}}{2}} \,,
\end{equation}
where the argument of $\tanh$ in \eqref{beta} vanishes. 
So, as $T$ increases $\l_*$ decreases. 
When $T$ becomes $T_c = \frac{\l}{\sqrt{2}}$, $\l_* = T_c$ 
the dilaton is perfectly screened by the black hole horizon. 
(i.e. If $T=0$,  $\l_* = \l$. If $T=\frac{\l}{\sqrt{2}}$, $\l_* = \frac{\l}{\sqrt{2}}$).
The point of the first order transition is where
the horizon screens the dilaton.

When density is introduced the story can become more complex. The action is (\ref{lagfin}) where it can be seen
from the first of the two terms in the square root that
including $d$ increases the action. This increase can be beneficial though if the second term
with $\beta$ can be reduced. It is possible to reduce the $\beta$ term if the D7 enters
the region where $\beta$ is large at small $\rho$. This means that the situation can arise where 
curving off the axis and then spiking on to the axis can be the lowest action state. This is typically
more likely where $\beta$ is largest in the interior space and the most savings
can be made entering that region at low $\rho$.  As we have seen at large values of $A$ embeddings
that spike onto the horizon do play a role introducing an extra phase.

It is only possible to have second order transitions if all three phases we have described are present.
In the D7 description the D7 must move from a curved embedding that avoids the black hole to a
configuration that spikes onto the black hole to a flat embedding smoothly.

\subsection{Dependence on the speed of running}

The parameter $\Gamma$ controls the period in $\rho$ or RG scale over which the change in the 
coupling $A$ occurs. It allows us to naively go from a precociously running theory to a walking
theory (although the change in the parameter $A$ over that period may enter into what is meant
by walking versus running too). 

\begin{figure}[]
\centering
\subfigure[$\G=0.1$  ]   
   {\includegraphics[width=5cm]{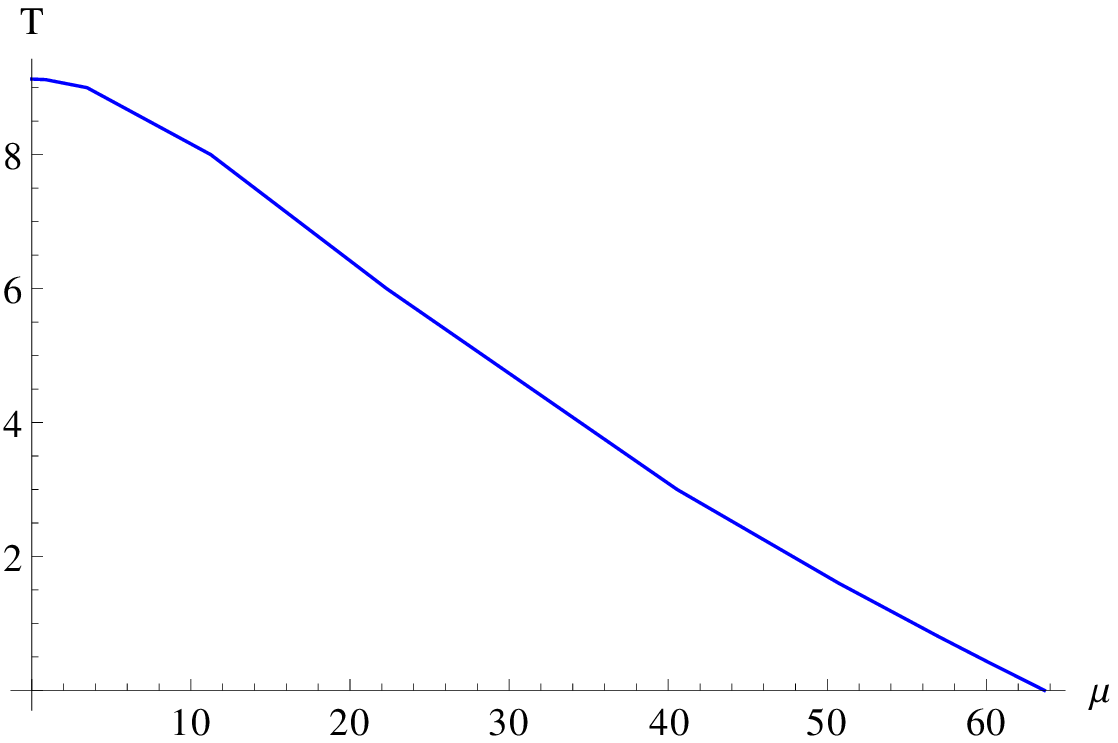}}
\subfigure[$\G=0.5$  ]   
   {\includegraphics[width=5cm]{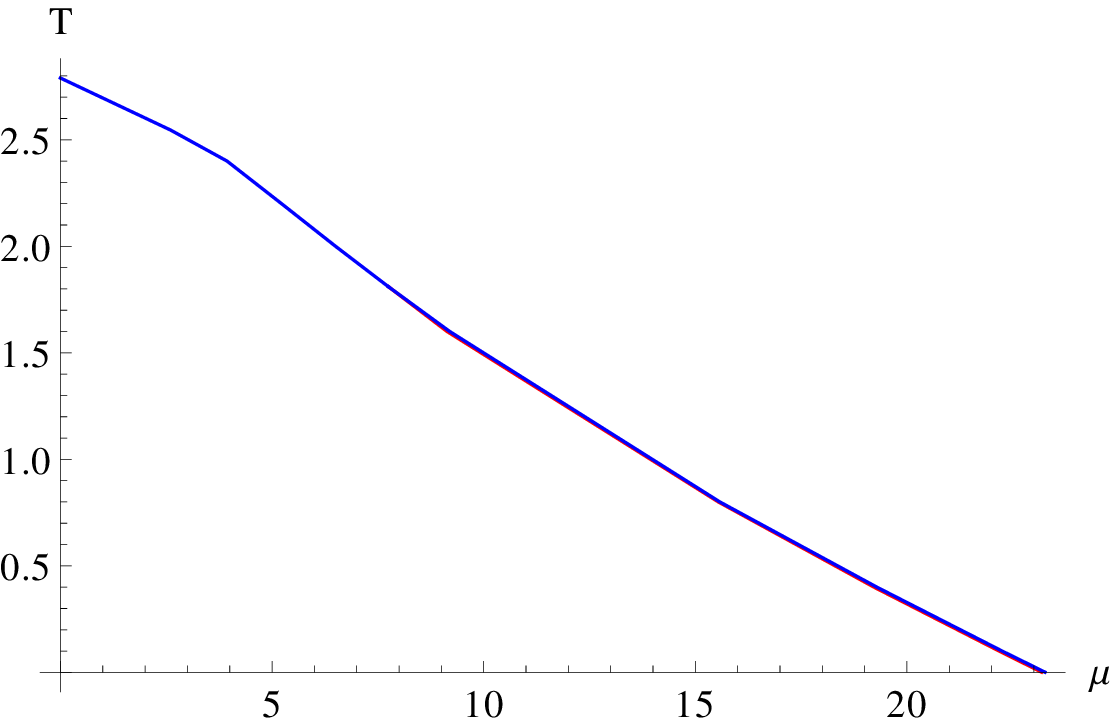}}
\subfigure[$\G=1$  ]   
   {\includegraphics[width=5cm]{fig6c30new.eps}}
  \caption{ Example plots of three possible phase structures for $A=30, \lambda = 1.715$
  and varying $\Gamma$.
  Large (small) $\G$ gives a second (first) order transition
    at low $T$.
    }
  \label{fig.Gamma}
\end{figure}

In Fig \ref{fig.Gamma} we show the phase diagram as a function of $\Gamma$ at fixed $\lambda$ and $A$.
We start at $\Gamma=1$ with a configuration already discussed that has all three phases present
and second order transitions at high density. As $\Gamma$ is reduced so that the step function in the dilaton
becomes broader the first order nature of the transitions reasserts itself. By $\Gamma=0.1$ the
mixed phase with chiral symmetry breaking but melted mesons is no longer preferred at any temperature
or chemical potential value - there is a single first order transition.  

In conclusion then moving towards a walking theory by either increasing the width of the running
or decreasing the magnitude of the increase in the coupling both move us towards a first order
chiral transition.  Stronger or quicker running favours a second order transition at low temperature, high density.

\section{Breaking the $\rho$-$L$ symmetry}

\begin{figure}[]
\centering
\subfigure[$\alpha=1, A=3,\G=1, \lambda=1.715$ ]
  {\includegraphics[width=5cm]{fig6a3new.eps}}
\subfigure[$\alpha=2.2, A=3,\G=1, \l=1.715$ ]
  {\includegraphics[width=5cm]{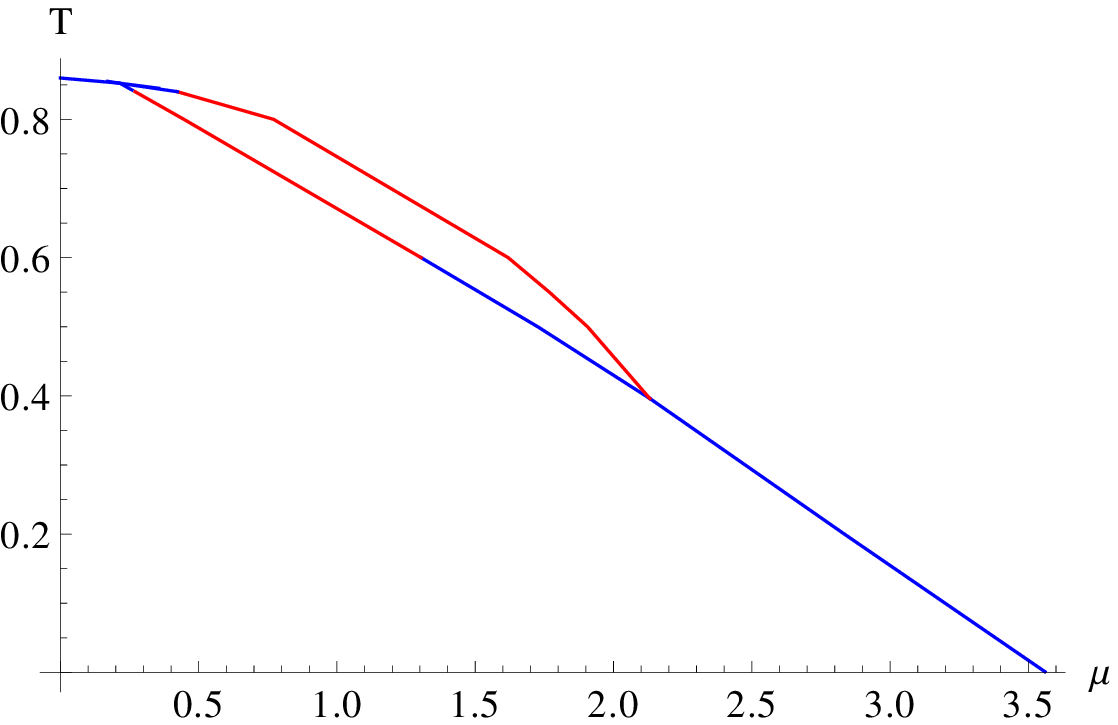}}
\subfigure[$\alpha=3, A=3, \G =1, \l=1.715$ ]  
   {\includegraphics[width=5cm]{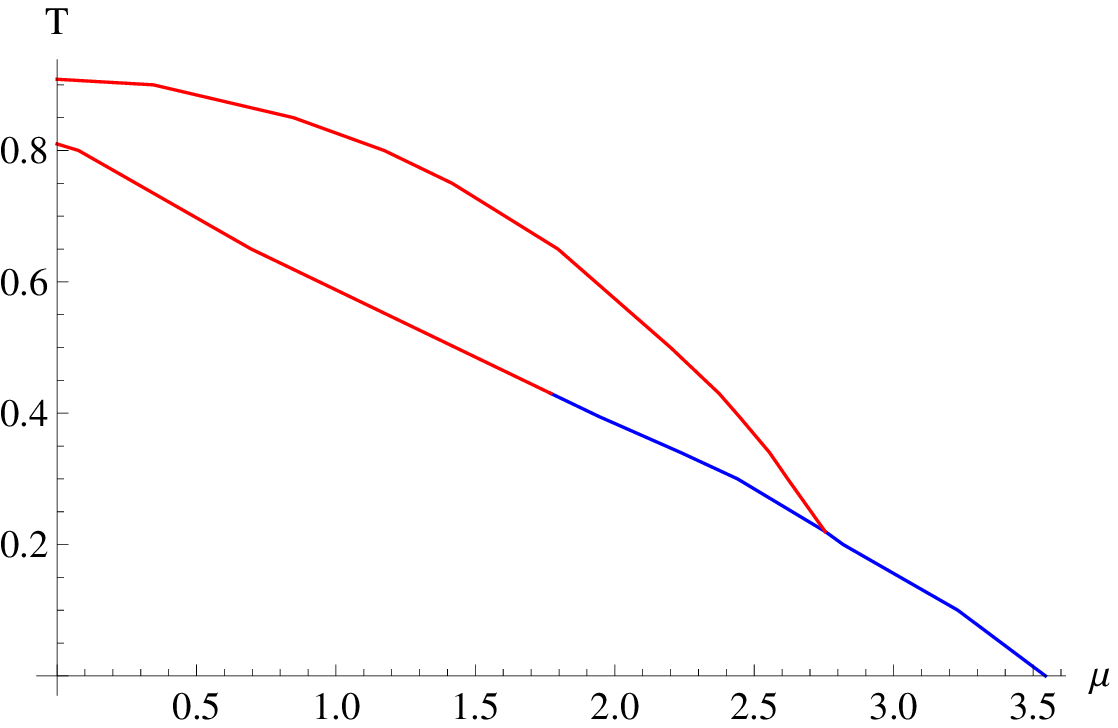}}   
  \caption{Sample phase diagrams for theories with none zero $\alpha$.  }\label{fig.elliptic2}
\end{figure}

Our goal is to attempt to reproduce a phase diagram comparable to that of QCD in our holographic model.
So far we have failed to generate a second order transition with temperature at zero density 
which is a key part of the QCD picture. 

We have a further natural freedom within our holographic model to exploit though. Our running dilaton
is in someway supposed to represent the backreaction of the quark fields on the strongly
interacting gauge dynamics to allow us to model theories with more interesting dynamics than the 
conformal ${\cal N}=4$ gauge fields. We introduce quarks  through D7 branes that break the 
$SO(6)$ symmetry of the five sphere of the original AdS/CFT Correspondence down to $SO(4)$ $\times$ $SO(2)$.
Our metric and ansatz for the running coupling (\ref{coup}) though respected the full $SO(6)$ symmetry. 
It seems reasonable to 
make use of the broken symmetry to introduce a further free parameter into our model. 

The most succesful scenario we have found is to  introduce our explicit $L-\rho$ symmetry breaking parameter, 
$\alpha$ through the blackening factors of the metric 
\begin{equation}
g_t = \frac{(w^4 - w_H^4)^2}{ w^4 (w^4+w_H^4)} \,,  \qquad
g_x  = \frac{w^4 + w_H^4}{ w^4} \,,
\end{equation}
with
\begin{equation}
  w^4 \ra \rho^2 + \frac{1}{\alpha} L^2 \,, \quad \alpha > 1 .
\end{equation}

We show the $\alpha$ dependence of our model in Fig \ref{fig.elliptic2}. We start from a model with a first order 
transition throughout the phase plane. As we increase $\alpha$ a region with melted mesons but chiral symmetry breaking
develops, associated with second order transitions. It then spreads to the zero chemical potential axis

The reason for the onset of second order transitions with just temperature is simply understood. We have deformed the
black hole horizon into an ellipse whose major axis is along the $L$ axis. The area of enlarged dilaton remains circular
in the $\rho$-$L$ plane. Thus there are temperature periods in which the area of the $\rho$-$L$ plane
with a large dilaton is covered except for a small piece that emerges from the horizon near the $\rho$ axis. 
If the value of the dilaton is sufficiently large in that uncovered area to encourage the D7 to
avoid it, but the horizon on the $L$ axis has met the zero temperature D7 embedding, then a second
order transition to a black hole embedding is likely. Since in the absence of the rise in the dilaton the
flat embedding would now be preferred the D7 settles on the horizon so it just misses the raised dilaton area.
As the black hole grows further the embedding is likely to track down onto the axis smoothly as the 
raised dilaton area is finally eaten. This intuition is indeed matched by the solutions as shown
in Fig \ref{fig.elliptic2}. 

The bottom phase diagram in Fig \ref{fig.elliptic2} achieves our goal of reproducing a 
chiral transition that is second order with temperature but first order with density.

\section{Summary}

In this paper we have converted the D3 probe-D7 system, that holographically describes ${\cal N}=4$ super
Yang-Mills theory with quenched ${\cal N}=2$ quark multiplets, to a phenomenological description
of strongly coupled quark matter. We introduced a simple unback-reacted profile for the dilaton that describes a
step of variable height and width in the running coupling of the gauge theory - (\ref{coup}). This breaks the conformal
symmetry of the model and introduces chiral symmetry breaking. We have then studied the temperature
and chemical potential phase structure of the model. 

The phase diagrams consist of three phases - a chirally symmetric phase at large temperature and 
density; a chirally broken phase with non-zero quark density at intermediate values of T and $\mu$;
and a chiral symmetry broken phase with zero quark density at low T and $\mu$. Fig \ref{Tvsmu}
shows these phase and their holographic analogue in a previously studied case where chiral 
symmetry was broken by an applied magnetic field. 
Here we showed that a small wide step in the gauge coupling's running gives rise to a single first order transition  between the chiral symmetric and the broken phase (see Fig \ref{fig.a}). If the step is made larger in height or thinner then the chirally broken phase with non-zero density also plays a role. Here the transitions at low temperature with chemical potential can be second order.
These results match known results in top down models in the presence of magnetic fields to induce the
symmetry breaking. 

We were interested in reproducing phase diagrams with the structure believed to exist in QCD. To do
this we made use of the broken $SO(6)$ symmetry of the gravity dual in the presence of D7 branes. Were
the branes backreacted the dilaton and geometry would reflect this symmetry breaking. We introduced a further 
phenomenological parameters $\alpha$ in the black hole blackening factor. This models allowed us to control which volumes of the holographic space have a large dilaton value
within, which the D7 branes prefer to avoid. Using this one extra parameter we were able to generate phase diagrams
like those in QCD with a chiral restoration transition
that was second order with temperature but first order with density (see Fig \ref{fig.elliptic2}). 

The ease with which such a variety of phase structures could be obtained is very encouraging for the
idea of phenomenologically modeling the QCD phase diagram holographically. Further the phenomenological
parameters we introduced are very natural in this context and it seems likely that top down
models with such phase structures should be possible as a result.

\acknowledgements NE is grateful for the support of
an STFC rolling grant.  AG, KK, and MM are grateful for University of Southampton Scholarships.  KK also acknowledge support via an NWO Vici grant of K. Skenderis  and University of Southampton Scholarships. This work is part of the research program of the Stichting voor Fundamenteel Onderzoek der Materie (FOM), which is financially supported by the Nederlandse Organisatie voor Wetenschappelijk Onderzoek (NWO).

\end{document}